\documentclass{ws-ijmpa}

\begin{document}

\markboth{B.S.Zou}{Multi-quark components in baryons}

\def\draftnote{}

\title{Multi-quark components in baryons}

\author{B.S. Zou}
\address{Institute of High Energy Physics, CAS, P.O.Box 918, Beijing
100049, China ;\\
Institute of Theoretical Physics, CAS, Beijing 100080, China}



\maketitle


\begin{abstract}
A brief review on some recent progresses in our understanding of
multi-quark components in baryons is presented. The multi-quark
components in baryons seem to be mainly in colored quark cluster
configurations rather than in ``meson cloud'' configurations or in
the form of a sea of quark-antiquark pairs. The colored quark
cluster multi-quark picture gives a natural explanation of
empirical indications for a positive strangeness magnetic moment
$\mu_s$ of the proton and the longstanding mass-reverse problem of
$S11(1535)$ and $P11(1440)$ $N^*$ resonances. A model-prediction
for the $\mu_s$ of the proton is given.
\end{abstract}



\section{Introduction}

In the classical quark model invented 40 years ago, a baryon is
composed of three quarks. The simple 3q constituent quark model
has been very successful in explaining the static properties, such
as mass and magnetic moment, of the ground states of the flavor
SU(3) octet and decuplet baryons. However, with advent of the QCD
in 1973 and later electron-proton deep inelastic scattering
experiments, we know that besides the three valence quarks in the
proton there are many other partons including quark-antiquark
pairs and gluons.

Before 1990, in all global analyses of parton distribution in
nucleons, a symmetric light-quark ($\bar u$, $\bar d$) sea was
assumed, based on the usual assumption that the sea of
quark-antiquark pairs is produced perturbatively from gluon
splitting \cite{Garvey}. However, a surprisingly large asymmetry
between the $\bar u$ and $\bar d$ sea quark distributions in the
proton has been observed in more recent deep inelastic scattering
\cite{NMC,HERMES} and Drell-Yan experiments \cite{NA51,NuSea}.

There have been many theoretical attempts
\cite{Kumano,Thomas,Nikolaev,Alberg,zyj} trying to find the
origins for this asymmetry. It is believed \cite{Garvey,NuSea}
that the asymmetry cannot be produced from perturbative QCD and
mesonic degrees of freedom play an important role for the effect.
In the popular meson-cloud model \cite{Thomas}, the excess of
$\bar d$ over $\bar u$ in the proton is explained by a mixture of
$n\pi^+$ with the $\pi^+$ composed of $u\bar d$. On the other
hand, the excess of $\bar d$ over $\bar u$ can be also explained
by a simple statistical model \cite{zyj,henley} by taking proton
as an ensemble of quark-gluon Fock states and using the principle
of detailed balance for transitions between various Fock states
through creation or annihilation of partons. The basic idea in
Ref.\cite{zyj} is rather simple:  while sea quark-antiquark pairs
are produced flavor blindly by gluon splitting, $\bar u$ quarks
have larger probability to annihilate than $\bar d$ quarks due to
the fact that there are more u quarks than d quarks in the proton,
which hence causes the asymmetry.

To understand the multi-quark components of the nucleon, the
strangeness in the proton is of particular interest, because it
definitely comes from the multi-quark part. In the next section,
we give a brief review of some recent progress on this aspect and
a simple model-prediction for the strangeness magnetic moment
$\mu_s$ of the proton. In section 3, we discuss how the inclusion
of penta-quark components can explain the longstanding
mass-reverse problem of $S11(1535)$ and $P11(1440)$ $N^*$
resonances and some relevant properties. Finally we give
conclusions in section 4.

\section{Strangeness in the proton}

Several measurements including the $\pi N$ $\sigma$-term,
neutrino-induced charm production and polarization effects in
electron-nucleon deep-inelastic scattering indicate that there may
be significant $s\bar s$ component in the proton
\cite{Ellis,Alberico,Beck}. Recently four experiments on parity
violation in electron-proton scattering suggest that the
strangeness magnetic moment of the proton $\mu_s$ is positive
\cite{mus-exp}. This is in contradiction with most theoretical
calculations. The meson-cloud calculations \cite{musolf} with a
fluctuation of the proton into a kaon and a strange hyperon lead
to a negative value for the strangeness magnetic moment $\mu_s$.
The statistical model \cite{zyj} gives a zero value for the
$\mu_s$. Various lattice QCD calculations give various values for
$\mu_s$ : $-0.28\pm 0.10$ \cite{Mathur:2000cf}, $+0.05 \pm 0.06$
\cite{Lewis:2002ix}, $-0.046 \pm 0.019\ \mu_{_N}$
\cite{leinweber}.

Very recently, a complete analysis \cite{zr} of the relation
between the $\mu_s$ and the possible configurations of the
$uuds\bar s$ component of the proton concludes that for a positive
$\mu_s$ value the $\bar s$ is in the ground state and the $uuds$
system in the $P$-state. The conventional $K^+\Lambda$
configuration has the $\bar s$ mainly in $P$-state and hence leads
to a negative $\mu_s$ value. The hidden strangeness analogues of
recently proposed diquark cluster models \cite{jaffe,zahed} for
the $\theta^+$ pentaquark have $\bar s$ in the ground state and
the $uuds$ system in the $P$-state, hence give a positive $\mu_s$
value. The analysis suggests that the $qqqq\bar q$ components in
baryons may be mainly in colored quark cluster configurations
rather than in the conventional ``meson cloud'' configurations.

In order to explain the observed excess of $\bar d$ over $\bar u$
and the non-zero $s\bar s$ contributions, the minimal
model-wave-function for the proton should be
\begin{equation}
|p> = N \{ |uud> + \epsilon_1 |[ud][ud]\bar d> + \epsilon_2
|[ud][us]\bar s> \}
\end{equation}
where $N=1/\sqrt{1+\epsilon_1^2+\epsilon_2^2}$, $[ud]$ and $[us]$
are scalar diquarks as in Jaffe-Wilczek diquark model
\cite{jaffe}. In order to reproduce the experimental measured
values $\bar d - \bar u \approx 0.12$ \cite{NuSea} and $2\bar
s/(\bar u + \bar d) \approx 0.48$ \cite{Baz95}, we have
$(N\epsilon_1)^2\approx 0.12$ and $(N\epsilon_2)^2\approx 0.03$.
Note here $\epsilon_2^2/\epsilon_1^2\approx 1/4$ which is smaller
than the corresponding SU(3) symmetry value 1/2 \cite{liu} due to
heavier $s$ quark mass than $u$, $d$ quarks. For this simple
model-wave-function, the probability of the $s\bar s$ component is
$P_{s\bar s}=(N\epsilon_2)^2\approx 0.03$, and the corresponding
strangeness magnetic moment $\mu_s$ can be calculated as \cite{zr}
\begin{equation}
\mu_s/\mu_{_N} ={m_p\over 3 m_s}(1+{2m_s\over m_{ud}+m_{us}})
P_{s\bar s} \, ,
\end{equation}
where $m_{ud}$ and $m_{us}$ are diquark masses for $[ud]$ and
$[us]$, respectively. Taking $m_u=m_d=360$MeV, $m_s=500$MeV,
$m_{ud}=720$MeV and $m_{us}=860$MeV as in Ref.\cite{liu}, we have
a prediction $\mu_s=0.043\mu_{_N}$. Since we have assumed zero
$uudu\bar u$ component in the simple model-wave-function, the
value should be regarded as the lower limit for the $\mu_s$. For
$\mu_s^{exp}$ the result of the SAMPLE experiment \cite{mus-exp}
is $\mu_s=(0.37\pm 0.34) \mu_{_N}$. More precise measurement of
the $\mu_s$ are crucial for finally pinning down the multi-quark
components in the proton.

\section{Penta-quark components in $N^*$ resonances}

From the study of the proton, we know that the probability of
multi-quark components in the proton are larger than 15\%. Then
for the excited nucleon states, $N^*$ resonances, there should be
more multi-quark components. To understand the properties of the
$N^*$ resonances, it is absolutely necessary to consider these
multi-quark components. A long-standing problem in conventional
$3q$ constituent quark models is to explain the mass reverse of
the $S11(1535)$ and $P11(1440)$ $N^*$ resonances.

Recently from BES results on $J/\psi\to\bar pp\eta$ \cite{ppeta}
and $\psi\to\bar pK^+\Lambda$ \cite{Yanghx}, the ratio between
effective coupling constants of $S11(1535)$ to $K\Lambda$ and
$p\eta$ is deduced to be $g_{N^*(1535)K\Lambda}/g_{N^*(1535)p\eta}
=1.3\pm 0.3$ \cite{lbc}. With previous known value of
$g_{N^*(1535)p\eta}$, the obtained new value of
$g_{N^*(1535)K\Lambda}$ is shown to reproduce recent $pp\to
pK^+\Lambda$ near-threshold cross section data as well. Taking
into account this large $N^*K\Lambda$ coupling in the coupled
channel Breit-Wigner formula for the $S11(1535)$, its Breit-Wigner
mass is found to be around 1400 MeV, much smaller than previous
value of about 1535 MeV obtained without including its coupling to
$K\Lambda$.

The nearly degenerate mass for the $S11(1535)$ and the $P11(1440)$
$N^*$ resonances can be easily understood by considering
multi-quark components in them. The $N^*(1535) 1/2^-$ could be the
lowest $L=1$ orbital excited $|uud>$ state with a large admixture
of $|[ud][us]\bar s>$ pentaquark component having $[ud]$, $[us]$
and $\bar s$ in the ground state.  The $N^*(1440)$ could be the
lowest radial excited $|uud>$ state with a large admixture of
$|[ud][ud]\bar d>$ pentaquark component having two $[ud]$ diquarks
in the relative P-wave. While the lowest $L=1$ orbital excited
$|uud>$ state should have a mass lower than the lowest radial
excited $|uud>$ state, the $|[ud][us]\bar s>$ pentaquark component
has a higher mass than $|[ud][ud]\bar d>$ pentaquark component.
The large mixture of the $|[ud][us]\bar s>$ pentaquark component
in the $N^*(1535)$ may also explain naturally its large coupling
to the final states with strangeness.

\section{Conclusions}

The probability of multi-quark components in the proton is at
least 15\%. The empirical indications for a positive strangeness
magnetic moment $\mu_s$ of the proton suggest the multi-quark
components in baryons to be mainly in colored quark cluster
configurations rather than in ``meson cloud'' configurations or in
the form of a sea of quark-antiquark pairs. The $\mu_s$ of the
proton is predicted to be no less than 4.3\% of the total magnetic
moment of the proton. The colored quark cluster multi-quark
picture gives a natural explanation of the longstanding
mass-reverse problem of $S11(1535)$ and $P11(1440)$ $N^*$
resonances, and the large coupling of the $S11(1535)$ to the
channels with strangeness.


\section*{Acknowledgements}
I thank B.C.Liu, D.O.Riska and Y.J.Zhang for collaboration on
relevant issues. The work is partly supported by CAS Knowledge
Innovation Project (KJCX2-SW-N02) and the National Natural Science
Foundation of China.



\begin{thebibliography}{0}


\bibitem{Garvey} G.T.Garvey, J.C.Peng,  Prog. Part. Nucl. Phys.
{\bf 47}, 203 (2001), and references therein.
\bibitem{NMC} New Muon Collaboration, M.Arneodo {\it et al.},
Phys. Rev. D {\bf 50}, R1 (1994).
\bibitem{HERMES} HERMES Collaboration, K.Ackerstaff {\it et al.},
Phys. Rev. Lett. {\bf 81}, 5519 (1998).
\bibitem{NA51} NA51 Collaboration, A.Baldit {\it et al.},
Phys. Lett. B {\bf 332}, 224 (1994).
\bibitem{NuSea} FNAL E866/NuSea Collaboration,
R.S. Towell {\it et al.}, Phys. Rev. {\bf D 64}, 052002 (2001).
\bibitem{Kumano} S.Kumano, Phys. Rep. {\bf 303}, 183 (1998).
\bibitem{Thomas} J.P.Speth and A.W.Thomas, Adv. Nucl. Phys. {\bf
24}, 93 (1997).
\bibitem{Nikolaev} N.N.Nikolaev {\it et al.}, Phys. Rev. D {\bf
60}, 014004 (1999).
\bibitem{Alberg} M.Alberg, E.M.Henley and G.A.Miller, Phys. Lett. B
{\bf 471}, 396 (2000).
\bibitem{zyj} Y.J. Zhang, B.Zhang and B.Q.Ma, Phys. Lett.
{\bf B523}, 260 (2001); Y. J. Zhang, B. S. Zou and L. M. Yang,
Phys. Lett. {\bf B528}, 228 (2002).
\bibitem{henley} M.Alberg and E.M.Henley, Phys. Lett. {\bf B611}, 111
(2005).
\bibitem{Ellis} J.Ellis, Nucl. Phys. {\bf A684}, 53c (2001).
\bibitem{Alberico} W.M.Alberico {\it et al.}, Phys. Rep. {\bf
358}, 227 (2002).
\bibitem{Beck} D.H.Beck and R.D.McKeown, Ann. Rev. Nucl. Part. Sci.
{\bf 51}, 189 (2001).
\bibitem{mus-exp} D. T. Spayde et al., Phys. Lett. {\bf B583}, 79 (2004);
K. Aniol et al., Phys. Rev. {\bf C69}, 065501 (2004); F. Maas et
al., Phys. Rev. Lett. {\bf 94}, 152001 (2005); D.S. Armstrong et
al., Phys. Rev. Lett. 95, 092001 (2005).
\bibitem{musolf} M. Musolf and M. Burkhardt, Z. Phys. {\bf C61}, 433
 (1984); H. Forkel, F. S. Navarra
and M. Nielsen, Phys. Rev. {\bf C61}, 055206 (2000); L. Hannelius
and D. O. Riska,  Phys. Rev. {\bf C62} 045204 (2000);  X.S.Chen et
al., Phys. Rev. {\bf C70}, 015201 (2004).
\bibitem{Mathur:2000cf} S.~J.~Dong, {\it et al.},
Phys.\ Rev.\ {\bf D58}, 074504 (1998).
\bibitem{Lewis:2002ix}
R.~Lewis, W.~Wilcox and R.~M.~Woloshyn, Phys.\ Rev. {\bf D67},
013003 (2003)
\bibitem{leinweber} D.B. Leinweber {\sl et al.}, Phys. Rev. Lett. {\bf
94}, 212001 (2005).
\bibitem{Baz95} A.O.Bazarko {\sl et al.}, Z. Phys. {\bf C65}, 189
(1995).
\bibitem{zr} B. S. Zou and D. O. Riska, Phys. Rev. Lett.
{\bf 95}, 072001 (2005).
\bibitem{jaffe} R. L. Jaffe and F. Wilczek, Phys. Rev. Lett. {\bf 91},
232003 (2003)
\bibitem{zahed} E. Shuryak and I. Zahed, Phys. Lett. {\bf B589}, 21 (2004)
\bibitem{liu} Y.R. Liu et al., Phys. Rev. {\bf C69}, 035205 (2004)
\bibitem{ppeta} J.Z.Bai et al.,(BES Collaboration), Phys. Lett. {\bf B510}, 75 (2001).
\bibitem{Yanghx} H.X.Yang et al., (BES Collaboration),
Int. J. Mod. Phys. {\bf A20}, 1985 (2005).
\bibitem{lbc} B. C. Liu and B. S. Zou, preprint nucl-th/0503069.
\end{thebibliography}
\end{document}